\documentclass[twoside,11pt]{article}
\usepackage{amsmath,amsfonts,amssymb,verbatim,float,anysize,fancyvrb,graphicx,multicol,mathrsfs,mdframed}
\usepackage[utf8]{inputenc}
\usepackage[doublespacing]{setspace}
\usepackage[english]{babel}
\usepackage{algorithm}
\usepackage{algpseudocode}
\usepackage{multirow}
\RequirePackage[colorlinks,citecolor=blue,urlcolor=blue]{hyperref}
\usepackage[numbers]{natbib}
\usepackage[usenames,dvipsnames, svgnames]{xcolor}
\usepackage{authblk}
\usepackage{multicol}
\usepackage{times}
\usepackage{subfig}
\usepackage{graphicx}
\usepackage{graphicx}
\graphicspath{{figures/}}
\usepackage{booktabs}
\usepackage[font=small,labelfont=bf]{caption}
\usepackage{amsfonts, amsmath, amsthm, amssymb}
\usepackage{wrapfig}
\usepackage[margin=2cm]{geometry}
\definecolor{mygreen}{rgb}{0,0.6,0.5}
\definecolor{myblue}{rgb}{0,0.45,0.7}
\definecolor{myred}{rgb}{0.8,0.4,0}
\definecolor{mygray}{rgb}{.6,.6,.6}

\parskip = \baselineskip
\newlength{\normalparindent}
\AtBeginDocument{\setlength{\normalparindent}{\parindent}}

\renewcommand{\arraystretch}{0.7}

\begin{document}
\setlength\parindent{24pt}

\title{Approximate Bayesian Inference on Mechanisms of Network Growth and Evolution}

\author{Maxwell H Wang \thanks{email: maxwang@hsph.harvard.edu}}
\author{Till Hoffmann}
\author{Jukka-Pekka Onnela}
\affil{\small{Department of Biostatistics, Harvard University}}

\maketitle

\begin{abstract}
{Mechanistic models can provide an intuitive and interpretable explanation of network growth by specifying a set of generative rules. These rules can be defined by domain knowledge about real-world mechanisms governing network growth or may be designed to facilitate the appearance of certain network motifs. In the formation of real-world networks, multiple mechanisms may be simultaneously involved; it is then important to understand the relative contribution of each of these mechanisms. In this paper, we propose the use of a conditional density estimator, augmented with a graph neural network, to perform inference on a flexible mixture of network-forming mechanisms. This event-wise mixture-of-mechanisms model assigns mechanisms to each edge formation event rather than stipulating node-level mechanisms, thus allowing for an explanation of the network generation process, as well as the dynamic evolution of the network over time. We demonstrate that our approximate Bayesian approach  yields valid inferences for the relative weights of the mechanisms in our model, and we utilize this method to investigate the mechanisms behind the formation of a variety of real-world networks.}
\end{abstract}

\section{Introduction}
Networks are a flexible tool for capturing the structure of non-random, heterogeneous connections among individuals in a population. By describing a population as a network of individuals, one can investigate the spread of behaviors and contagious diseases, and evaluate how social interactions facilitate or prevent their spread. For such purposes, it is critical to understand the mechanisms behind the formation and evolution of these networks.

Mechanistic network models seek to provide an intuitive and interpretable explanation of network growth by specifying a set of generative rules that govern network growth. These rules can be defined by domain knowledge about real-world mechanisms governing the appearance of edges, or they may be chosen to facilitate the appearance of certain network motifs. A well-known example of a mechanistic network model is the Barabási–Albert (BA) model \citep{albert2002}, in which the network grows one node at a time and edges are added via a linear Preferential Attachment mechanism. More complex models have sought to add the effects of additional factors such as node fitness \citep{bianconi2001} and temporal changes in attachment probabilities \citep{wang2008}, or to use local neighbor information rather than global information to determine the addition of new edges \citep{davidsen2002, mossa2002, vazquez2003}. Other models, such as duplication-divergence models, seek to model the growth behavior of gene regulatory and protein interaction networks \citep{bhan2002, teichmann2004, ispolatov2005}, while agent-based models seek to replicate the behavior of independent agents each seeking to maximize some set of goals \citep{snijders1996, snijders2010, niezink2019}.

The genesis of real-world social networks can often be the product of multiple mechanisms operating simultaneously. For example, a  network may have nodes that exhibit both Preferential Attachment behavior and random attachment behavior. Previous work in this field has proposed a mixture model of mechanisms for network growth processes \citep{ratmann2007, de2019mixed, langendorf2021, raynal2022} where, at each time step, a mechanism is sampled from a weight vector $\tilde{\alpha} = (\alpha_1,...,\alpha_M)$ and applied. The target of inference is then $\tilde{\alpha}$, which specifies the relative importance, or frequency of occurrence, of each mechanism. The number of edges attributed to each node may be described by a mechanism-dependent parameter. For example, one mechanism may be associated with Random Attachment, whereby a new node joins the network and connects to a fixed number of $m_{RA}$ nodes selected uniformly at random among the existing nodes. Here, $m_{RA}$ would be a constant value that must also be inferred on. 

To understand the processes that govern the formation of networks, as well as their relative importance, a significant body of work has aimed to conduct inferences or model selection for mechanistic network models. In the area of model selection, previous work has proposed bootstrapping methods \citep{chen2019bootstrap}, alternating decision trees \citep{janssen2012}, and the Super Learner algorithm \citep{chen2020flex}. For statistical inferences, certain reversible mechanistic models allow for likelihood-based approaches \citep{larson2023}. Other methods have sought to bridge the gap between mechanistic and statistical models, which can define a likelihood for all possible networks, through novel statistical models such as the congruence class model (CCM) \citep{goyal2023}. However, mixture-of-mechanisms models do not generally have tractable likelihoods. For many real-world networks, only the ``final" network is observed once all nodes and edges have been added to the network; intermediate stages of network growth are not observed, and thus the order with which nodes and edges were added to the network is typically unknown. In addition, certain mechanisms, such as in duplication-divergence models, may remove edges or nodes, thus erasing evidence of the actions of previous mechanisms. Indeed, the inability to describe an exact likelihood across possible networks is one of the primary drawbacks of utilizing mechanistic network models over statistical network models such as exponential random graph models (ERGMs).

Due to the unavailability of a tractable likelihood, statistical inferences on mixture-of-mechanisms models require the use of approximate inference methods, such as approximate Bayesian computation (ABC) and Mixture Density Estimation \citep{bishop1994}. ABC denotes a set of methods first introduced in the study of population genetics \citep{tavare1997}. These methods seek to create an approximation of the posterior distribution through rejection sampling. Proposal parameter values are sampled from the prior and used to generate simulated data. If the simulated data are deemed ``similar enough" to the observed data, the proposal parameter value is accepted as a sample from the approximate posterior distribution. 

ABC methods typically require the manual specification of a relatively small set of summary statistics, and their performance typically degrades if too many summaries are used due to the curse of dimensionality \citep{blum2010}. When the likelihood is within the exponential family, Bayes-sufficient summary statistics will be available. In these cases, if the width of the acceptance kernel is set to zero, ABC methods can be made exact, in the sense that $P(\theta|S(X)) = P(\theta|X)$. Here, $\theta$ is the set of parameters to be inferred (such as the weight vector $\tilde{\alpha}$), $X$ is the observed data, and $S(X)$ is the vector of summary statistics calculated from $X$.

However, most models that call for the use of ABC, including most mechanistic network models, do not have Bayes-sufficient summary statistics and require the user to provide a set of summaries deemed to be informative about the mechanisms. Previous studies investigating mixtures of mechanisms tended to use \emph{ad hoc} summary statistics, including clustering coefficient, motif counts, or degree distribution \citep{ratmann2007, langendorf2021}, but it is not apparent that these preselected summary statistics encompass all informative summary statistics or cannot represented with a lower-dimensional transformation.

Work closely related to ours proposes the use of the Random Forest ABC \citep{raynal2019, raynal2022} to select the summary statistics with the best ABC predictive performance. In this method, a large pool of candidate network statistics is initially considered. These statistics include metrics on degree distribution (e.g., mean degree, degree entropy, maximum degree), as well as counts of certain network motifs (e.g., triangles, four-cliques). The statistics are then iteratively ranked by ABC predictive performance, and the best $T$ statistics are chosen for use in the ABC procedure, where $T$ is a value chosen by the user (in \citep{raynal2022}, $T=10$). This method can perform well, especially when paired with algorithms such as SMC-ABC \citep{toni2009} and provides an intuitive understanding on which summary statistics are most informative for inferences on the relative weights of mechanisms. However, the Random Forest ABC is subject to many of the same drawbacks as other summary statistic selection methods: it must have an adequately informative pool of initial statistics to select from, and it cannot discover more informative, lower-dimensional, transformations of the initial statistics. Furthermore, some of these network summary statistics are very computationally expensive; while it is possible to lessen the computational load by subsampling from each simulated network and computing the statistics for the network sample \citep{raynal2020scalable}, the potential loss of information that this subsampling may cause has not yet been fully investigated.

Instead of utilizing manually specified network statistics for inference, we aim to extract informative summary statistics from the network in an automated manner. This problem of network information extraction is closely related to whole-graph embedding, where a network is mapped onto a lower dimensional embedding space. There exists a broad variety of methods for whole-graph embedding, including matrix factorization methods \citep{wang2019} and kernel methods \citep{kriege2020}. Other methods are based on deep learning, such as the Graph2Vec framework \citep{narayanan2017} and graph autoencoders \citep{tian2014, wang2016}.

We propose to conduct conditional density estimation \citep{bishop1994} by augmenting a simple Mixture Density Network (MDN) with a graph neural network (GNN). Here, despite the similar network-related terminology, the Mixture Density Network is distinct from the mixture-of-mechanisms network model we proposed earlier. Instead, it serves as a tool for learning informative summary statistics. Mixture Density Networks were first proposed by \citep{bishop1994} for modelling conditional probability distributions, and were later used for learning parameterized approximation to the posterior distribution in Bayesian contexts \citep{papamakarios2016}. While many applications of MDNs tend to utilize simple multilayer perceptrons, this would be inefficient when inferring on potentially sparse network data. Furthermore, a dense network architecture would be sensitive to the network size, and due to being a non-permutation-invariant architecture, results would be sensitive to different orderings of nodes \citep{lee2019}. Instead, we will utilize graph neural networks (GNNs), first introduced by \citep{gori2005}, which learn a permutation invariant transformation of the attributes of the nodes and edges of a graph. By utilizing a GNN in series with an MDN, we can efficiently extract network information that is then used to learn a parameterized approximation of the posterior distribution of the parameters governing the mixture-of-mechanisms model of network growth. 

In this paper, we will motivate and describe a novel event-based version of the mixture-of-mechanisms growth model which applies mechanisms at the edge level and allows for changes in nodes already present in the network. We will then discuss the architecture of the proposed GNN-MDN method, and demonstrate and validate the usage of the method on simulated datasets. Finally, we will apply the GNN-MDN to infer on the relative weights of mechanisms within several real-world social networks.

\section{Model}

In this paper, we propose an adjusted version of existing mixture-of-mechanisms models. In previous mixture-of-mechanisms models, mechanisms are typically associated with a node, and all edges associated with the node make use of the same mechanism (e.g., if a node is associated with the Preferential Attachment mechanism, all of its edges will be attached via Preferential Attachment). Thus, mechanisms are always assigned at the node level. Our approach is motivated by the observation that, for example, in many social networks, individual nodes do not necessarily correspond to a single network mechanism. For example, consider a model with two mechanisms: Random Attachment and Preferential Attachment. If we consider a new user in a social media network, if mechanisms are assigned at the node level, the new user would either attach solely to random nodes (e.g., low-degree personal friends) or primarily to high-degree nodes (e.g., popular celebrities). 

\begin{figure}
\begin{center}
\includegraphics[width=6in]{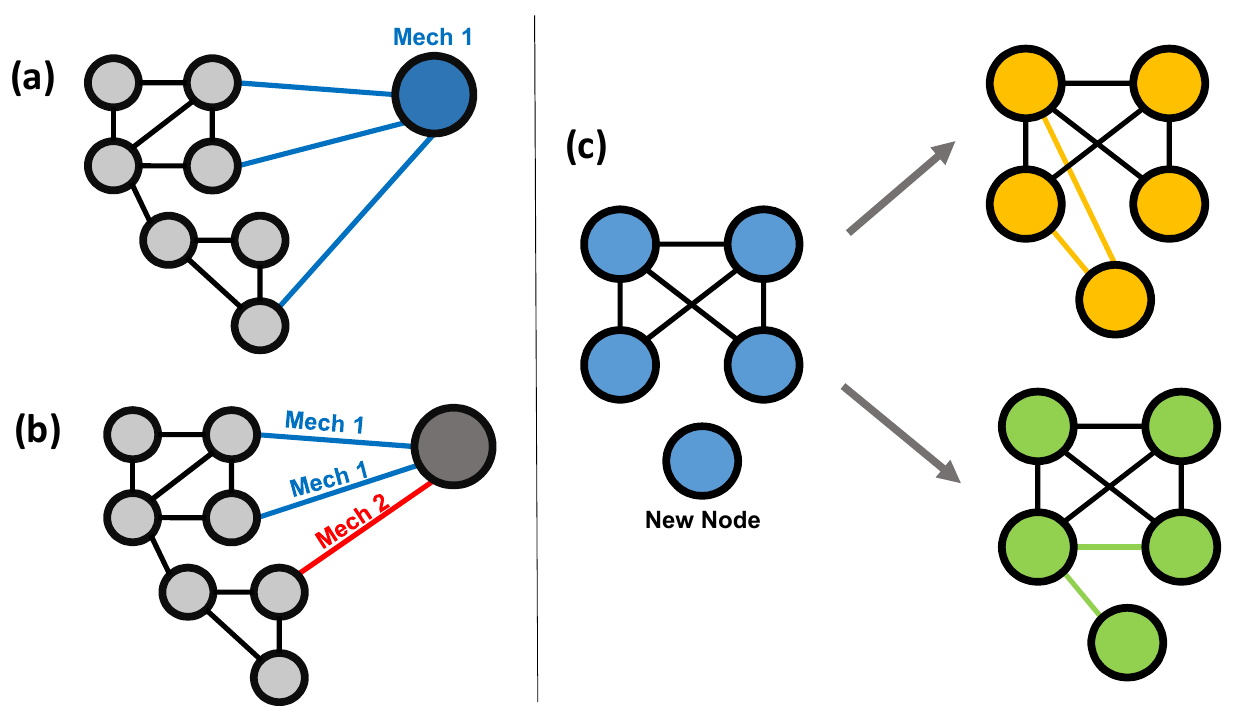}
\caption{In a nodewise mixture-of-mechanisms model \textbf{(a)}, mechanisms are assigned to each incoming node. In contrast, in an edgewise mixture-of-mechanisms model \textbf{(b)}, each edge can be assigned to a different mechanism. In \textbf{(c)}, a new node with 2 edges is to be added to the network, such that the number of triangles in the network is to be maximized. Under most growth models, edges can only be added between the new node and the existing network (top). By allowing for new edges between existing nodes, we can create one additional triangle (bottom), for a total of four triangles in this network} \label{explanation}
\end{center}
\end{figure}

In many real-world networks, it is likely that individuals instead form multiple connections based on different mechanisms \citep{meyer1994, hernandez2020}. In the social media example, a new user could potentially form connections with both popular celebrities (via Preferential Attachment) and their relatively lesser-known personal friends (via cyclic closure \citep{holme2002, kumpula2009}). Thus, mechanisms are associated with the addition of edges in the network rather than the nodes. While our edge-based model is similar to models in the literature \citep{holme2002, dorogovtsev2000, gomez2006}, it allows for a wider range of mechanisms and greater flexibility in the way mechanisms are applied to events. 

In addition, many mechanistic models are defined solely in the context of network growth, where edges are added and removed purely in relation to newly added nodes. Two well-known examples of this are the  Barabási–Albert (BA) model \citep{albert2002} and the Duplication-Divergence model \citep{ispolatov2005}. However, this restriction limits the flexibility of potential network structures. Thus, we propose the inclusion of evolutionary mechanisms that act on existing edges \emph{without} adding nodes. This flexibility is important for modeling features like triangle formation or degree assortativity. For example, when seeking to maximize the number of triangles in a network, one would prefer to complete as many cliques as possible, which may require adding edges to existing nodes.

To formalize our model, we define $N$ growth mechanisms $\{m_{g1}, ..., m_{gN}\}$ and $M$ evolution mechanisms $\{m_{e1},...,m_{eM}\}$. Growth mechanisms are assigned a weight $\alpha_i \in [0,1]$, such that $\sum_{i=1}^N \alpha_i = 1$. Evolution mechanisms are assigned a weight $\beta_i \in [0,1]$, such that $\sum_{i=1}^M \beta_i = 1$. Now, simulation of the network is as given in Algorithm \ref{mm_algo}.

\begin{algorithm}
\caption{Edgewise Mixture-of-Mechanisms Network Formation Model}
\begin{algorithmic}
\State Begin with seed network with $c_0$ nodes. The current total number of nodes is $c \geq c_0$, and the network is to be grown to $c_f$ nodes.
\While {$c < c_f$}
    \State Add node $c$ to network.
    \State Draw number of growth events $v_g \sim P_g(v_g|\phi_g)$.
    \ForAll {$i \in \{1, ...,v_g\}$}
        \State Draw growth mechanism $m_g \in \{m_{g1}, ..., m_{gN}\}$ based on weights $\alpha$.
        \State Apply mechanism $m_g$.
    
    \EndFor
    \State Draw number of evolution events $v_e \sim P_e(v_e|\phi_e)$.
    \ForAll {$i \in \{1,...,v_e\}$}
        \State Draw growth mechanism $m_e \in \{m_{e1}, ..., m_{eM}\}$ based on weights $\beta$.
        \State Apply mechanism $m_e$.
    \EndFor
\EndWhile
\end{algorithmic}
\label{mm_algo}
\end{algorithm}

Here, we introduced the random variables $v_g \sim P_g(v_g|\phi_g)$ and $v_e \sim P_e(v_e|\phi_e)$ to denote the number of events associated with the addition of a single new node. The number of growth events, $v_g$, is associated with the degree of the new node. While application of traditional mechanistic models, such as the BA model, sometimes make the assumption that all new nodes are added with the same number of nodes, by treating the degree of incoming nodes as a random variable, we allow some degree of flexibility in the degree of incoming nodes. This idea has existed since early work by \cite{price1976}. Note that both of these probability distributions, $P_g(v_g|\phi_g)$ and $P_v(v_e|\phi_e)$, include the parameters $\phi_g$ and $\phi_e$, which must also be inferred. In this paper, we will assume that $P_g$ and $P_e$ take the form of Poisson distributions, similar to a previously described Poisson growth model for networks \citep{sheridan2008}. In this case $\phi_g = \lambda_g$ and $\phi_e = \lambda_e$, where $\lambda_g$ is the rate of the distribution $P_g$ and $\lambda_e$ is the rate of the distribution $P_e$.

The likelihood for the edgewise mixture-of-mechanisms model is not generally tractable. In the following section, we will discuss methods that may be used to infer the parameters $(\phi_g, 
\phi_e, \tilde{\alpha}, \tilde{\beta})$.


\section{Method}

\subsection{GNN-MDN}

In Bayesian inference, Mixture Density Networks (MDN) \citep{bishop1994, papamakarios2016} methods seek to learn a parameterized mixture density that approximates the true posterior distribution, typically through the use of a feed-forward neural network. The outputs of the neural network density estimates of the parameters; for example, if the  posterior distribution is to be approximated by a mixture of two independent Gaussian distributions, the neural network would learn the means and variances of the two Gaussian components and their relative weights in the mixture. 

While the adjacency matrix of the network could potentially be used as the input for an MDN, this is inefficient in both memory use and computation time. A multilayer perceptron (MLP), the typical architecture for MDNs, does not inherently utilize the network structure within the data, though an MLP, given sufficient depth and width, can still theoretically learn features that pertain to the structure of the network. Furthermore, most realistic networks are sparse, such that the bulk of the adjacency matrix for each simulated network provides little additional information. 

Instead, it is preferable to extract relevant information directly from the network data. This goal interfaces closely with the rich existing literature on whole-graph embedding, where critical information describing the high-dimensional graph is summarized with some lower-dimensional embedding. There exists a wide range of approaches for embedding, including matrix factorization \citep{wang2019}, kernel-based methods \citep{kriege2020}, language-processing methods \citep{narayanan2017}, and graph autoencoders \citep{tian2014, wang2016}. Many of these methods were developed in the context of graph classification and clustering, where the embedding was chosen to minimize some classification error. However, in our setting, the goal is Bayesian inference rather than classification. Thus, instead of minimizing a classification error, we aim to minimize the Expected Posterior Entropy (EPE) \citep{hoffmann2022}. As discussed in \citep{hoffmann2022}, minimizing the EPE is equivalent to learning an approximate posterior distribution that minimizes the Kullback-Leibler Divergence between the approximate posterior and the true posterior distribution. The EPE is also the loss function that is minimized by MDNs.

\begin{figure}
\begin{center}
\includegraphics[width=5in]{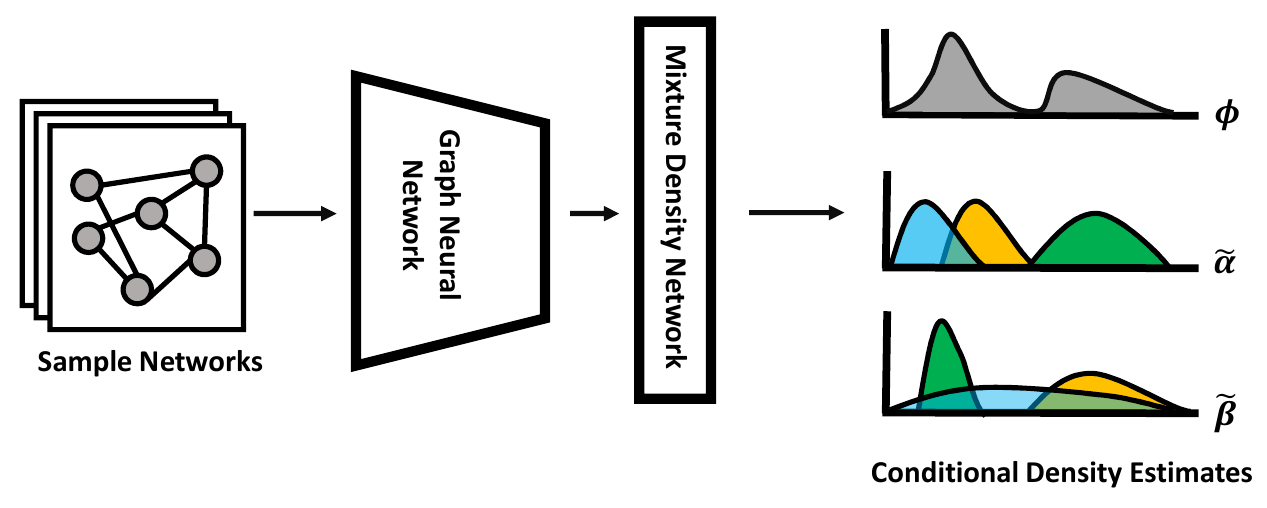}
\caption[GNN-MDN Schematic]{A Graph Neural Network (GNN) and a Multilayer Perceptron (MLP), which serves as our Mixture Density Network (MDN), are trained in series to minimize Expected Posterior Entropy (EPE). The resultant output is a conditional density estimate, which approximates the posterior distribution of parameters $\phi_g$, $\phi_e$, $\tilde{\alpha}$, and $\tilde{\beta}$.} \label{gnn_mdn_schematic}
\end{center}
\end{figure}

For the purpose of inferring the parameters of our edgewise mixture-of-mechanisms model, we will utilize a Graph Neural Network (GNN) in series with a traditional feed-forward MLP to learn the mixture density estimates. GNNs were first developed by \cite{gori2005} and are commonly used for applications in image classification \citep{xu2017, yang2018} and whole-graph embedding \citep{hamilton2017}. A wide range of GNNs exist, including spectral-based methods, such as the Graph Convolutional Network (GCN) \citep{kipf2016}, spatial-based methods, such as the Graph Attention Network (GAT), \citep{velivckovic2017} and the Graph Isomorphism Network (GIN) \citep{xu2018}. We will primarily utilize the Higher-Order GNN proposed by \citep{morris2019}, though other architectures may be considered. The GNN will be followed by a simple MLP, serving as our MDN, which is trained to minimize the EPE. The output of the MDN will then be a conditional density estimate of our parameters of interest. A schematic of this approach is shown in Figure \ref{gnn_mdn_schematic}. We will refer to the GNN-based MDN as the GNN-MDN. Finally, while a GNN-based approach can be  extended to similar ABC methods such as MDN-ABC \citep{hoffmann2022}, we will focus primarily on conditional density estimation.

\section{Simulation Study}

To demonstrate the utility of GNN-MDN for inferences on relative weights of mechanisms, we will simulate networks using known weight vectors and examine the results. Furthermore, we will show that the GNN-MDN method fulfills the coverage property, which is a necessary condition for valid approximate Bayesian inference \citep{talts2018}.

In the following, we define mechanisms for network growth. Let $i$ be the node that was most recently added to the network, and let $\mathcal{N}_i$ be the set of nodes that were in the network before node $i$ was added. Let $d_j(i)$ be the degree of node $j$ at the time node $i$ is added. The simulated networks consisted of 400 nodes, and each was started from a seed network of 4 nodes. Our network growth mechanisms are:

\begin{enumerate}
    \item \textbf{Random Attachment} (RA): An edge is added between node $i$ and a node selected uniformly at random from $\mathcal{N}_i$.
    \item \textbf{Preferential Attachment} (PA): An edge is added between node $i$ and node $j$, where node $j$ is selected from $\mathcal{N}_i$ with probability $\frac{d_j(i)}{\sum_{k \in \mathcal{N}_i} d_k(i)}$. This mechanism was first introduced by \citep{albert2002}.
    \item \textbf{Negative Preferential Attachment} (NPA): An edge is added between node $i$ and node $j$, where node $j$ is selected from $\mathcal{N}_i$ with probability $\frac{(d_j(i) + \epsilon)^{-1}}{\sum_{k \in \mathcal{N}_i} (d_k(i) + \epsilon)^{-1}}$. Here, $\epsilon$ is a small constant that accounts for nodes with zero degree; we set $\epsilon = 10^{-4}$. This mechanism is similar to that proposed by \citep{pollner2005}, though in that reference, the attachment probability scales as $e^{-d_j(i)}$.
\end{enumerate}

In addition, we define the mechanisms for network evolution. While not necessary, we constrained each of these mechanisms to preserve the total number of edges in the network (i.e., for every edge added to the network, one is removed). Our network evolution mechanisms are:

\begin{enumerate}
    \item \textbf{Triangle Formation} (TF): Find a node $j$ that has neighbors $k$ and $l$. Add an edge between $k$ and $l$, and remove a random edge from the network such that the edge is not currently part of a triangle. The first part of this mechanism behaves similarly to the one described by \citep{davidsen2002}.
    \item \textbf{Disassortative Rewiring} (NPA-PA): Select node $j$ with probability $\frac{(d_j(i) + \epsilon)^{-1}}{\sum_{k \in \mathcal{N}_i} (d_k(i) + \epsilon)^{-1}}$, node $k$ with probability $\frac{d_k(i)}{\sum_{l \in \mathcal{N}_i} d_l(i)}$, and add an edge between nodes $j$ and $k$. Remove a random edge. This mechanism decreases the assortativity \citep{newman2002assortative} of the network.
    \item \textbf{Preferential Rewiring} (RA-PA): Select node $j$ uniformly at random, select node $k$ with probability $\frac{d_k(i)}{\sum_{l \in \mathcal{N}_i} d_l(i)}$, and add an edge between nodes $j$ and $k$. Remove a random edge.
\end{enumerate}

To determine the number of growth and evolution events drawn at each time step, we define $v_g \sim \mbox{Poisson}(\lambda_g)$ and $v_e \sim \mbox{Poisson}(\lambda_e)$, where $\lambda_g$ and $\lambda_e$ are both parameters that must be inferred on. Thus, the targets of inference are $\theta = (\lambda_g, \lambda_e, \tilde{\alpha}, \tilde{\beta})$. We defined the priors as $\lambda_g \sim \mbox{Gamma}(2,2)$, $\lambda_e \sim \mbox{Gamma}(2,2)$, $\tilde{\alpha}\sim\mbox{Dirichlet}(0.5,0.5,0.5)$, and $\tilde{\beta}\sim\mbox{Dirichlet}(0.5,0.5,0.5)$. In this simulation section, the true values of each parameter are shown in Table \ref{gnn_mdn_scenario_table}. For these scenarios, $\tilde{\beta}$ values were chosen to emulate situations where one primary mechanism is dominant over the others. When applied to empirical data, all of these parameters are unknown and would need to be inferred. Priors were selected to include reasonable probability mass at the ``true" values of parameters while still being relatively disperse. For each scenario, the networks were generated to have 500 nodes, which is on the same order of magnitude as the real-world networks explored in the Data Application section.

\begin{table}[!h]
\begin{center}
\caption{Settings of parameters for simulation scenarios.}
\label{gnn_mdn_scenario_table}
\renewcommand{\arraystretch}{1.5} 
\begin{tabular}{ |c|c|c|c|c| }
\hline
Scenario & $\lambda_g$ & $\lambda_e$ & $\tilde{\alpha}$ & $\tilde{\beta}$\\
\hline
1 & 5 & 5 & (0.95, 0.025, 0.025) & (0.025, 0.95, 0.025) \\ 
2 & 5 & 5 & (0.025, 0.95, 0.025) & (0.95, 0.025, 0.025) \\ 
3 & 5 & 5 & (0.025, 0.95, 0.025) & (0.025, 0.025, 0.95)\\
\hline
\end{tabular}
\end{center}
\end{table}

To conduct inferences on the mechanisms governing the formation of some observed network $G_0$, the edge list of network $G_0$ is introduced to the GNN-MDN, and the resultant mixture density estimate can be interpreted as $P(\theta|G_0) = P(\lambda_g, \lambda_e, \tilde{\alpha}, \tilde{\beta}|G_0)$.

\subsection{MDN Settings}

To train the MDN, we utilized 500,000 simulations in the training set and 250,000 simulations in the validation set. To extract the network information, we utilized a GNN with five layers. The GNN was trained to learn 20 features per node; these features were then mean-pooled across each network and fed into an MLP with 3 hidden layers. The overall MDN learned five components, with each component being the independent combination of two Dirichlet distributions (describing the weight vectors $\tilde{\alpha}$ and $\tilde{\beta}$) and two Gamma distributions (describing the Poisson rates $\lambda_g$ and $\lambda_e$). 

At each epoch, the loss (EPE) was computed for the validation set. Training continued until 10 epochs elapsed without improvement in validation loss, or until 100 total epochs elapsed. We utilized an Adam optimizer \citep{kingma2014} with a learning rate of $0.0005$.

We evaluated the minimal EPE attained on the validation dataset using three different GNN architectures: the Graph Convolutional Network (GCN) \citep{kipf2016}, the Higher-Order GNN \citep{morris2019}, and the Graph Isomorphism Network (GIN) \citep{xu2018}. These modules are adopted from the PyTorch Geometric library, and though they are all convolutional layers, they differ in the specific update rules. In addition, to compare with a non-GNN approach, we also learned 20 features using graph2Vec \citep{narayanan2017}, and fed those features into an MLP to obtain a conditional density estimate. This graph2vec approach learns graph embeddings based on the similarity of the rooted subgraphs of the networks included in the corpus, rather than seeking to minimize the EPE.

\begin{figure}
\begin{center}
\includegraphics[width=4in]{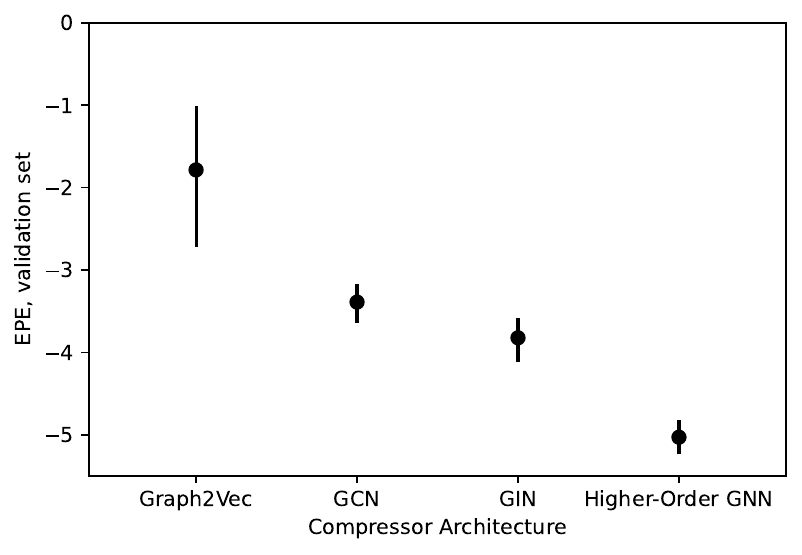}
\caption{Validation loss for four different architectures. EPEs are obtained as Monte Carlo estimates over the entire validation dataset. We obtained error bars by splitting the validation set into 250 sections; the maximum and minimum of the error bars correspond to the maximum and minimum EPE evaluated across those subsets.} \label{comparisons}
\end{center}
\end{figure}

The comparisons appear to indicate that the Higher-Order GNN is able to learn the best approximation (as measured by EPE) to the true posterior distribution; thus, in all of the following sections, we will utilize the Higher-Order GNN architecture for all of the GNN-MDN inferences. Graph2vec yields the least optimal summary statistics, as the graph embeddings used in this approach are learned without consideration to minimizing the EPE. Note, however, that each of these architectures involve hyperparameters (such as the input and output dimensions of each convolutional layer) that could potentially be more optimally tuned to obtain better performances.

\subsection{Simulation Results}

\begin{figure}
\begin{center}
\includegraphics[width=6in]{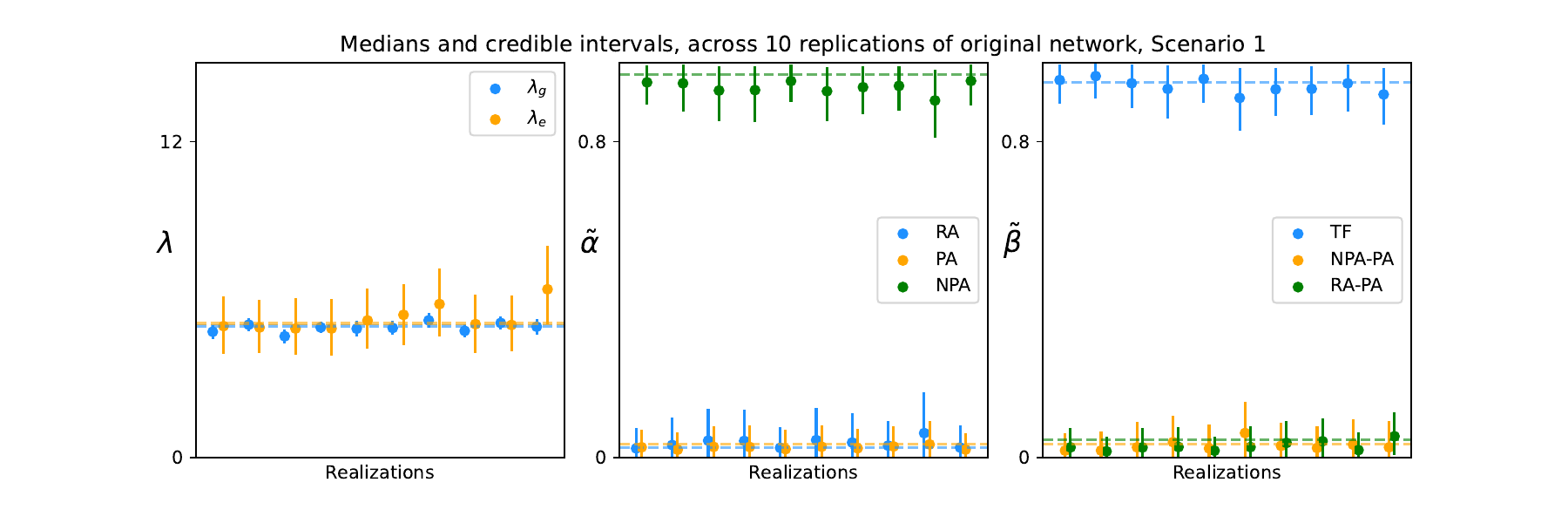}
\includegraphics[width=6in]{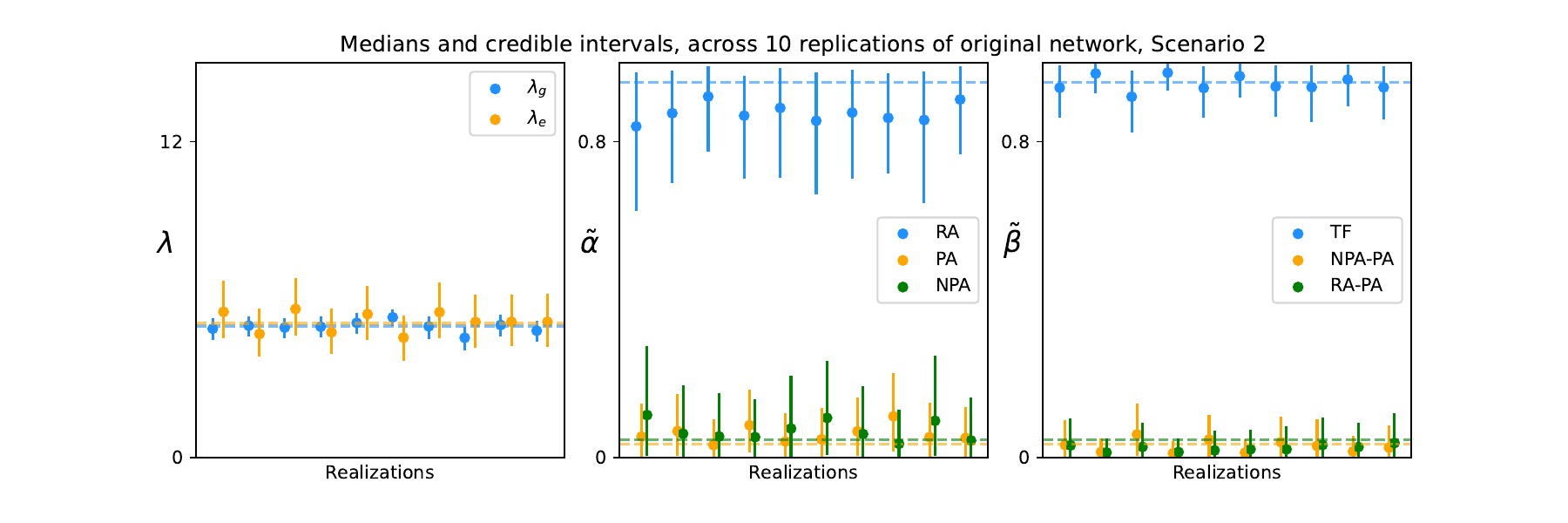}
\includegraphics[width=6in]{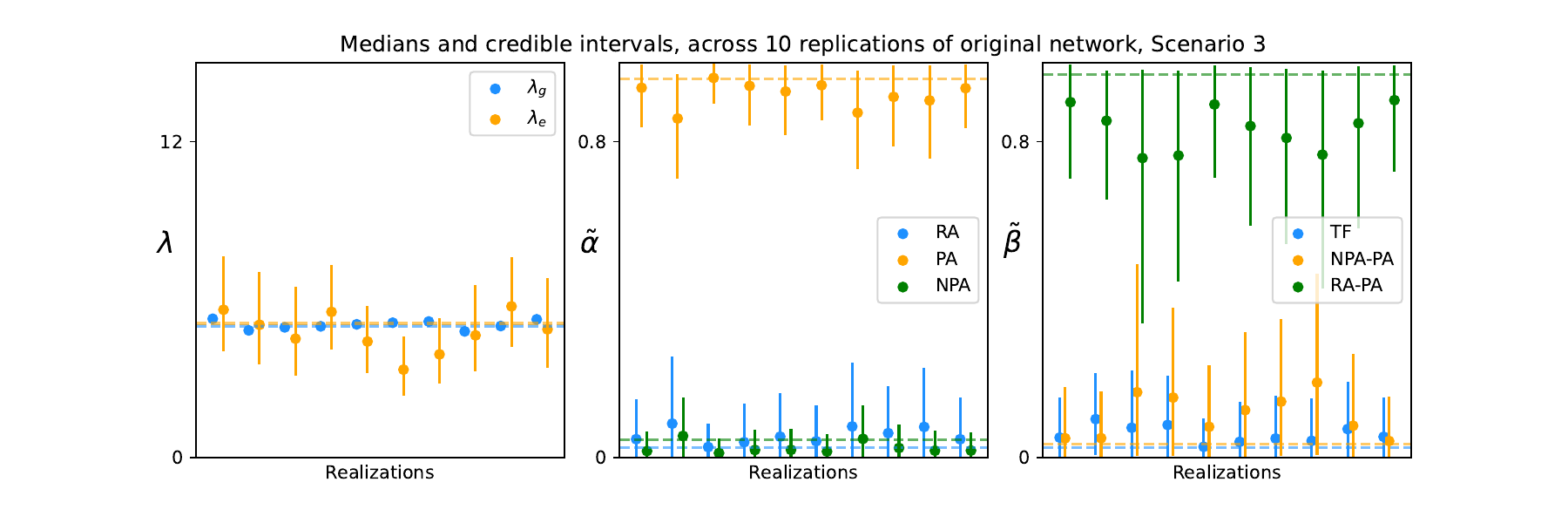}
\caption[Simulation Results, GNN-MDN]{Simulation results for three scenarios. For each scenario, we generated 10 realizations (replications) of the true network and utilized GNN-MDN to estimate the 95\% credible intervals and medians. True values for each parameter are shown as dotted horizontal lines. Due to the stochasticity of the data generation process behind the true networks, the credible intervals associated with each realization are expected to be different.} \label{gnn_mdn_sims}
\end{center}
\end{figure}

To demonstrate the use of the GNN-MDN method, we will consider three scenarios, each with different true values for the parameters of interest, shown in Table \ref{gnn_mdn_scenario_table}. For each scenario, we used the true values of the parameters to generate 10 realizations of an `observed network and obtained the approximate posteriors of the parameters of interest in Figure \ref{gnn_mdn_sims}. Notably, due to the inherent stochasticity of the network model, the posterior distributions associated with each set of parameters will vary across each replication.

Due to the large number of parameters in our model, we limited our analysis to only a small number of scenarios. In order to evaluate and validate the performance of the GNN-MDN method across a wider span of the parameter space, we will examine coverage properties in the following subsection. 

\subsection{Coverage Properties}

When conducting approximate Bayesian inference, it is crucial to evaluate the quality of the approximation. To do this, we evaluated the coverage properties of the credible intervals derived from the approximate posteriors \citep{cook2006, prangle2014_diagnose, talts2018}. We drew 1000 samples of $\theta' = (\lambda_g, \lambda_e, \tilde{\alpha}, \tilde{\beta})$ from their priors, and for each sample, we simulated a network $G'$ from the model. For each simulated network $G'$, we obtained posterior distributions $P(\lambda_g, \lambda_e, \tilde{\alpha}, \tilde{\beta}|G')$. For each marginal posterior distribution of parameter $\theta \in (\lambda_g, \lambda_e, \tilde{\alpha}, \tilde{\beta})$, we defined the $\gamma\%$ credible interval as $I = (I_l, I_u)$, such that $\mbox{Pr}(\theta\in I|Y_{obs})=\gamma/100$, and $\mbox{Pr}(\theta < I_{l}|Y_{obs}) = \mbox{Pr}(\theta > I_{u}|Y_{obs}) = \frac{(1-\frac{\gamma}{100})}{2}$. The empirical coverage of a $\gamma\%$ credible interval is then the percentage of times that the true value of a parameter falls within that interval. If the coverage property holds, the empirical coverage of a $\gamma\%$ credible interval should be approximately $\gamma\%$. As shown in \citep{prangle2014_diagnose}, significant departures from linearity would indicate that the approximate posterior distribution is a poor match to the true posterior.

\begin{figure}
\begin{center}
\includegraphics[width=6in]{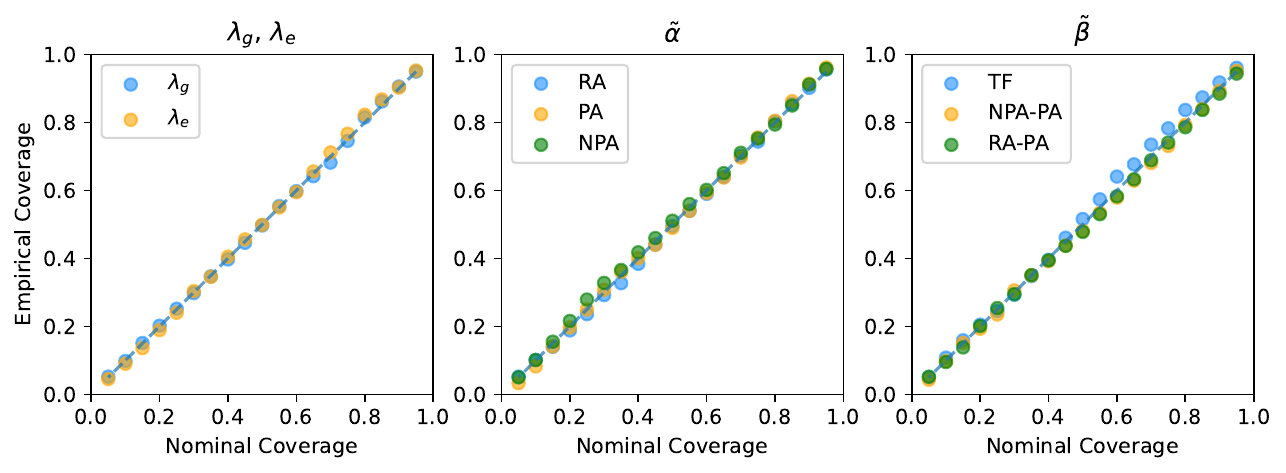}
\caption[GNN-MDN Coverage Properties]{Coverage properties for parameters of interest, using posterior samples of GNN-MDN. If coverage properties are fulfilled, all points should lie on the identity line.} \label{GNN_MDN_coverages}
\end{center}
\end{figure}

\begin{figure}
\begin{center}
\includegraphics[width=6in]{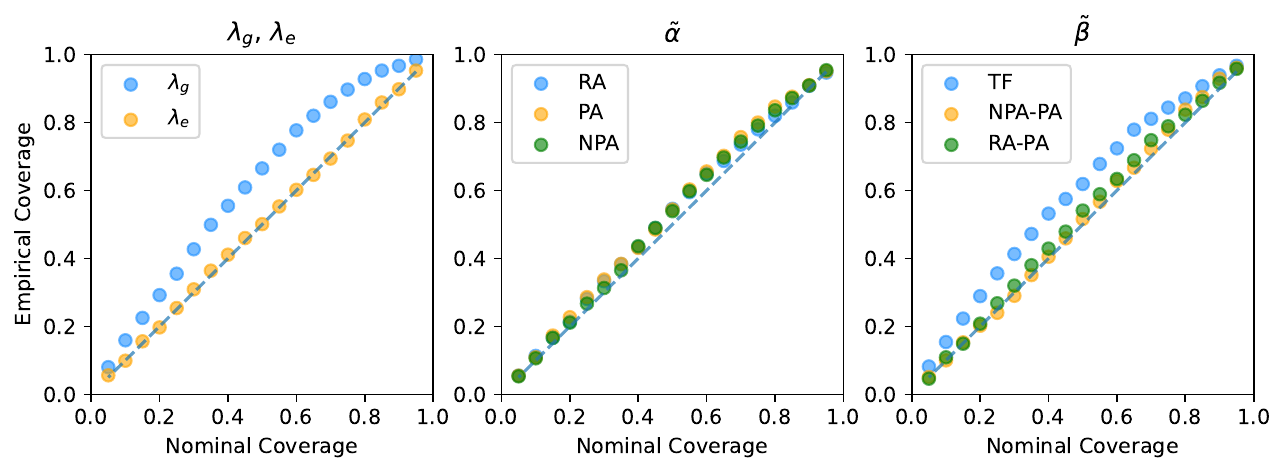}
\caption[Rejection ABC Coverage Properties]{Coverage properties for parameters of interest, using posterior samples from rejection ABC. If coverage properties are fulfilled, all points should lie on the identity line.} \label{rejection_abc_coverages}
\end{center}
\end{figure}

We also compared the coverage properties of the GNN-MDN method to a rejection ABC algorithm that utilizes network statistics at the summary statistics. To do this, we followed the procedure in \citep{raynal2022}. We generated a pool of 100,000 simulations, and for each iteration of ABC, we drew the best 0.2\% of samples (around 200 samples) to make up our ABC sample \citep{prangle2014_diagnose}. The distance metric was the Euclidean distance between a vector composed of 10 selected summary statistics from \citep{raynal2022}, including average degree and number of triangles. All summary statistics were normalized before L2 norms were computed. The resultant coverages, associated with approximate inferences based on network summary statistics, are shown in Figure \ref{rejection_abc_coverages}. Notably, while the the credible intervals produced by the GNN-MDN in Figure \ref{GNN_MDN_coverages} appear to fulfill the coverage property, simply using network summary statistics in rejection ABC leads to visible issues with coverage in Figure \ref{rejection_abc_coverages}.

\section{Data Application: Social Networks}

In this sectionm we will demonstrate the practical utility of our method by inferring model parameters for three real-world social networks, each collected through different means, where individuals are treated as nodes and the interactions between them are edges. Since we are utilizing the same mechanisms as the simulation study, all networks are treated as static, undirected, and unweighted.

The three networks we will demonstrate our method on are as follows:

\begin{enumerate}
    \item The Enron network, collected from email data.
    \item The SPHH Conference network, collected from proximity data.
    \item The Karnataka microfinance network, collected from survey data.
\end{enumerate}

The Enron network is based off the Enron Corpus \citep{klimt2004}, which contained the emails between 158 senior employees of the Enron corporation. We flattened it as an unweighted, undirected network, where any emails between two individuals is indicative of an edge.

The SPHH Conference network \citep{genois2018} is based on data collected through wearable sensors worn by participants in a scientific conference. Because contact data was collected every 20 seconds, the original dataset is a temporal network. We flattened this network by simply assuming that an unweighted edge exists between any two individuals with at least 6 recorded instances of proximity (corresponding roughly to around two minutes of total contact). This was meant to eliminate spurious edges that could have occurred due to brief, incidental encounters between individuals.

Finally, the Karnataka microfinance dataset was collected as part of a survey-based study where participants were asked to nominate social contacts based on a variety of interactions (giving of advice, borrowing of money, etc.). The publicly available dataset \citep{banerjee2013} contains both individual and household-level data from 77 Indian villages; we used the individual-level data from Village 10 only. We counted an edge as existing between two individuals if either of them nominated the other for \emph{any} of the survey questions. While this is essentially a flattened version of a union of networks, we note that in this dataset, the different types of interactions tend to exhibit a high degree of overlap.
\begin{figure}  
\includegraphics[width=\textwidth]{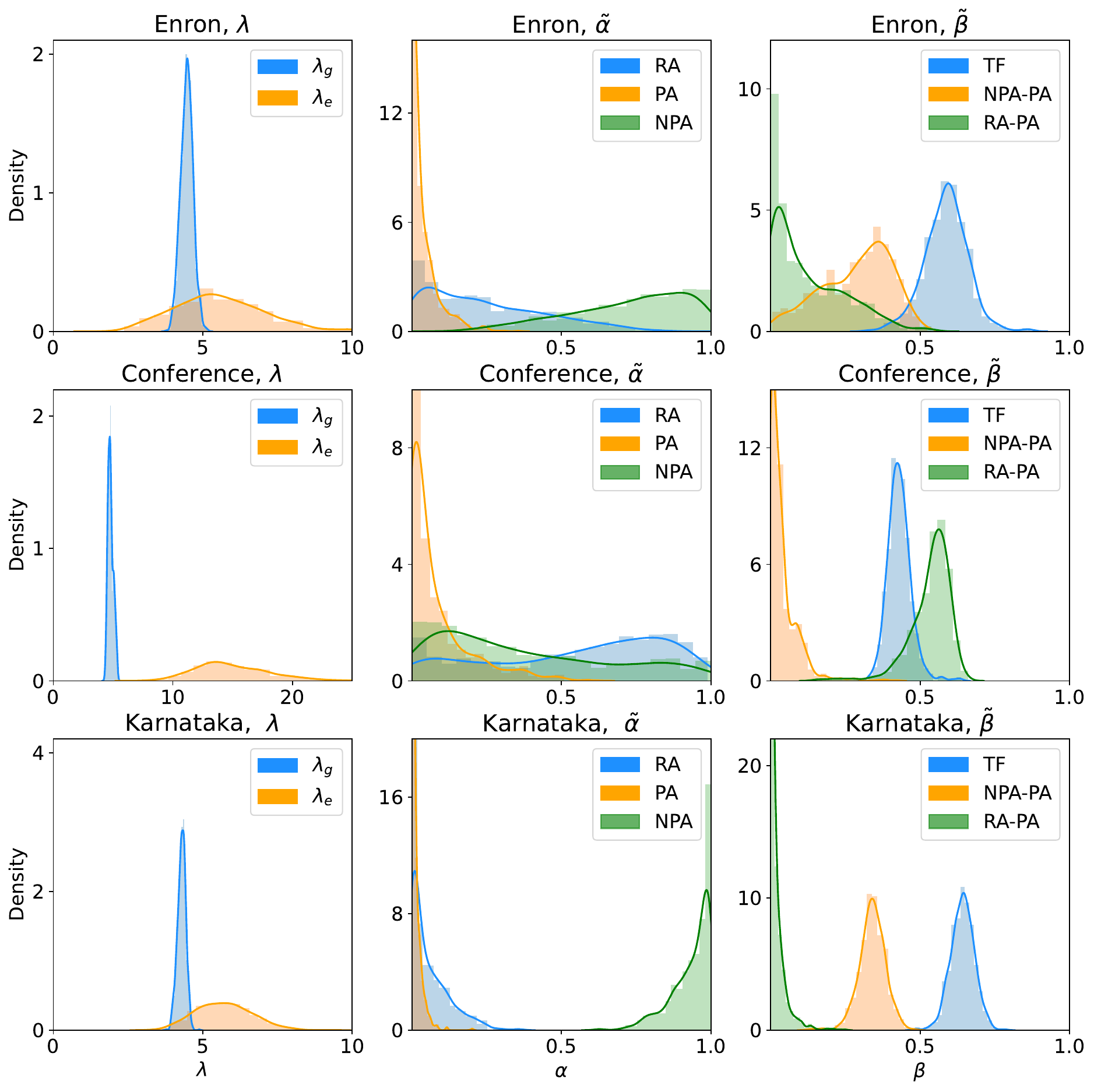}
\caption[Mechanism weights for networks of sociality]{Marginal posterior densities for parameters $\lambda_g$, $\lambda_e$, $\tilde{\alpha}$, and $\tilde{\beta}$ for observed networks: Enron, Conference, and Karnataka. 1000 samples were drawn for each parameter. Gaussian kernel density estimates of posterior samples are included for visual clarity.\label{combined_densities}}
\end{figure}

We conducted inference on the parameters $(\lambda_g, \lambda_e, \tilde{\alpha}, \tilde{\beta})$ using the GNN-MDN. All networks were trained using the same architecture as given in the simulation study and utilized the same priors. Approximate posterior densities are shown in Figure \ref{combined_densities}.

\begin{figure}
\begin{center}
\includegraphics[width=6in]{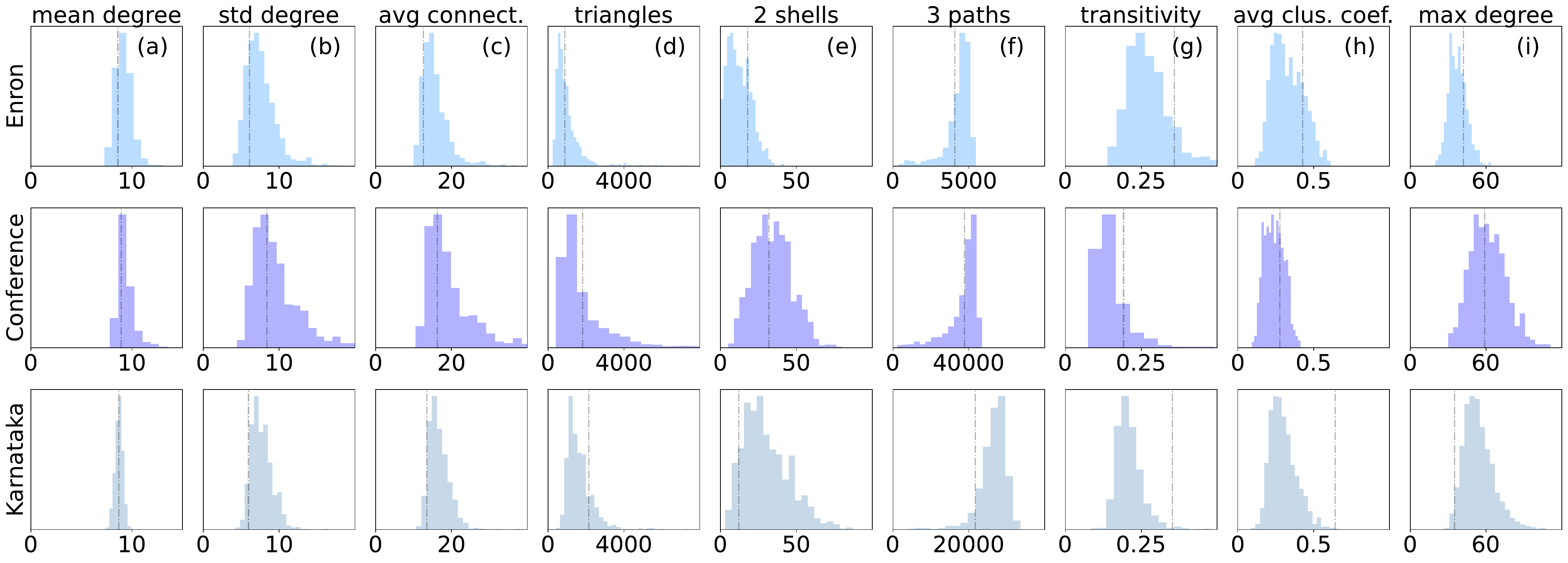}
\caption[Posterior predictive check of social networks]{Posterior predictive checks for a variety of network summary statistics, conducted by sampling values of $(\lambda_g, \lambda_e, \tilde{\alpha}, \tilde{\beta})$ from the posterior distribution, forward-sampling 1000 networks, and evaluating the network statistics. These network statistics are: a) mean degree, b) standard deviation of degree, c) degree entropy, d) number of triangles, e) number of 2-shells, f) number of 3-shortest paths, g) transitivity, h) average clustering coefficient, and i) maximum degree. True observed values of each statistic, calculated from the observed network, are plotted with a dotted vertical line} \label{network_pp_checks}
\end{center}
\end{figure}

Finally, we evaluated how well the model captured crucial features of these three observed networks. For each network, we drew 1000 posterior predictive samples and evaluated a series of network summary statistics shown in Figure \ref{network_pp_checks}. Essentially, we sampled $(\lambda_g, \lambda_e, \tilde{\alpha}, \tilde{\beta})$ from our estimated posterior distribution $P(\lambda_g, \lambda_e, \tilde{\alpha}, \tilde{\beta}|G)$, and used those parameters to forward-simulate 1000 networks from our network model. All of these simulated networks have the same number of nodes as the original network dataset they are based off of. For each of these 1000 networks, we sampled the value of our network summary statistics (e.g., the mean of the degree distribution). In general, if our mixture-of-mechanisms model is a good fit for these real networks, the posterior predictive distributions of these network summary statistics should include the true (unknown) values of these statistics, calculated from the observed network.

Notably, while the posterior predictive checks yielded reasonable results for the Enron and Conference networks, it appears that the high degree of clustering in the Karnataka network is not captured by our mechanistic model. One reason for this is that the Karnataka network tends to be separated into many disparate cliques, usually based on household and family membership. None of the six mechanisms we investigated were able to directly induce the formation of community structures. In order to rectify this discrepancy, mechanisms that take family membership or other covariates into account when forming edges or mechanisms that induce network communities could be added to the model.

\section{Discussion}

In this paper, we have investigated the use of the GNN-MDN to conduct posterior inferences on an edgewise mixture-of-mechanisms model. This model can be readily extended to include other mechanisms or be adjusted to account for missingness or noise in network observations \citep{young2020}. Unlike existing models where each node is assigned a mechanism, our model allows for mechanisms to be assigned at the edge level, which may be a more accurate representation of edge formation. While the mechanisms we chose are not specific to social networks, they  are general enough to be utilized across a variety of networks. We demonstrated our approach on several real-world networks, and our results highlight the importance of the triangle formation mechanism. Interestingly, none of the inferences attributed a high weight to the Preferential Attachment mechanism. However, to ascertain whether certain mechanisms play \emph{any} role in the formation of a network, one would have to extend this framework to model selection. The validity of ABC for model selection has been questioned \citep{robert2011}, but other approaches have been proposed for model selection for mechanistic models \citep{chen2020}.

Because the network mechanisms we investigated were meant to be  generalizable, we did not include more advanced mechanisms that could take into account factors such as node-level covariates or temporal trends in social behavior. However, there are many directions for the implementation of more advanced mechanisms. For instance, mechanisms could be designed for temporal networks, which could account for both recurring, long-term interactions and weaker, more transient interactions \citep{karsai2014, gelardi2021}. Other mechanisms could be designed to include node-level covariates; nodes exhibiting certain traits may be assumed to be more attractive than others (i.e., a fitness model), or nodes may seek to attach to nodes exhibiting similar covariates (i.e., homophily). For example, the Karnataka network consists primarily of tightly connected familial groups. In such cases, it may be necessary to introduce a mechanism that is able to capture the high level of clustering seen in the  network; for example, we may allow nodes to exhibit homophilous tendencies by preferentially attaching to other nodes that share node-level covariates (family membership, age, etc.).

The mixture-of-mechanisms model we utilized included six mechanisms, but these mechanisms are by no means exhaustive. In addition, while we focused on static, unweighted, undirected networks, this method can be generalized to temporal, weighted, or directed networks, as long as the mechanisms within the model are capable of generating networks comparable to the observed data. 

Beyond mechanistic network models, the combination of GNNs and MDN literature provides a new avenue for Bayesian inferences on a variety of processes where the output is a network. For example, similar to the framework explored by \citep{young2020}, there may exist situations where dyadic interaction data is observed, but missingness and noise are present in the data. Our approach could, at least in principle, be used to infer the mechanisms even in this setting as long as the resultant network observations can be simulated, given model parameters. While the approach in \citep{young2020} is most suitable when observations are independent at the dyadic level, the GNN-MDN approach requires no such assumption, as no tractable likelihood is required, and it might therefore generalize to more realistic settings.

Mechanistic models allow for a wide degree of flexibility in studying social and biological networks. In this paper, we presented an approximate Bayesian inference framework that can be utilized for mixtures-of-mechanisms models, but also other mechanistic models that can be forward-simulated given specific parameter values. While there remains significant work to be done to continue bridging the gap between statistical and mechanistic network models, the GNN-MDN method provides an approach to gain a deeper understanding of the various mechanisms that govern the structure of real-world networks.

\section{Acknowledgements}
This work was supported by the National Institutes of Health [T32AI007358, R01 AI138901].

\bibliography{references}

\end{document}